\input{aipcheck}

\documentclass[
    ,final            
  ]
  {aipproc}

\newcommand{\ee}{e^{+} e^{-}}

\newcommand{\ccbar}{c\bar{c}}
\newcommand{\qqbar}{q\bar{q}}
\newcommand{\ppbar}{p\bar{p}}
\newcommand{\leplep}{\ell^{+}\ell^{-}}
\newcommand{\jp}{J/\psi}
\newcommand{\psip}{\psi '}
\newcommand{\jpsi}{J/\psi}

\newcommand{\mumu}{\mu^{+}\mu^{-}}
\newcommand{\pipi}{\pi^{+}\pi^{-}}

\newcommand{\chiconep}{\chi_{c1}^{\prime}}

\newcommand{\rt}{\rightarrow}

\layoutstyle{8x11double}

\begin{document}

\title{ Recent Results from BaBar, Belle, BESIII and CDF }


\author{Stephen Lars Olsen}{
  address={Department of Physics \& Astronomy, Seoul National University, Seoul 151-747, KOREA}
}

%

\begin{abstract}
A brief report of some recent experimental developments concerning 
the $X$, $Y$ and $Z$ charmoniumlike mesons states and other puzzling
states from the BaBar, Belle, BESIII and CDF experiments is presented.
\end{abstract}

\maketitle


\section{Introduction}

The $XYZ$-mesons are an assortment of meson resonances
discovered by BaBar, Belle, BESIII and CDF
-- somewhat haphazardly named $X$, $Y$ or $Z$ -- that have defied
assignments to the quark-antiquark $\qqbar$ meson structure specified
by the classical quark-parton model (QPM).  Most of them are
seen to have decays to final states with a charmed-quark 
anticharmed-quark ($\ccbar$) pair, which almost certainly means
that they have a $\ccbar$ pair among its constituents.  However,
the spectrum of particles that are comprised of only a $\ccbar$ pair
-- the so-called charmonium mesons -- is very well understood, the number of
unassigned levels is small and the properties of whatever
fill them are tightly constrained. It is now generally
agreed that at least some of the newly discovered $XYZ$ mesons have a
 more complex substructure than the $\qqbar$ mesons of the QPM.
What, in fact, this more complex structure may be remains an
open question.  One peculiar feature that may be a clue
to their ultimate understanding, is that many of these new states have
partial decay widths for hadronic transitions 
to standard charmonium meson states -- such as the
$\jp$, the $\psip$ and the $\chi_{c1}$ -- that are much larger
than is typical for the established  $\ccbar$ mesons.

Other unusual states have been reported.  BESII
found a large enhancement ine the $\ppbar$ invariant mass spectrum
right at the $M(\ppbar )=2m_p$ mass threshold in radiative $\jp\rt\gamma\ppbar$
decays.  And the Belle group found a huge $\pipi \Upsilon (nS)$
($n=1,2~\&~3$) peak in the $\ee\rt\pipi\Upsilon (nS)$ cross-section
around 10.9~GeV.  It is not know if either of these are related to the
$XYZ$ mesons.

Although some of these phenomena have been around for a number of years
their origins have still not been identified.  This remains an experimentally
driven subject and the hope is that with enough information, the puzzle (puzzles?)
can be solved.  In this talk I briefly review some recent experimental results
that may have some relevance to their interpretation.

\section{The $X(3872)$}

The $X(3872)$ was discovered by Belle in 2003
as a narrow peak in the $\pipi\jp$ invariant mass distribution
from $B^+\rt K^+\pipi\jp$ decays~\cite{belle_x3872,conj}.
This peak was subsequently confirmed by
CDF~\cite{CDF_x3872}, D0~\cite{D0_x3872} and BaBar~\cite{babar_x3872}.
CDF and D0 see $X(3872)$ produced promptly in inclusive $p\bar{p}$ collisions as well
as in $B$ meson decays.   In all of the experiments, the invariant mass distribution of the
dipion system is consistent with originating from $\rho\rt\pipi$~\cite{CDF_pipi},  
indicating that the
$C$-parity of the $X(3872)$ is $C=+1$.  Charmonium states are all isospin singlets;
the decay charmonium$\rt\rho\jp$ violates isospin and should be strongly suppressed.  
A study of angular correlations among the $\pipi\jp$ final state particles by CDF
led to the conclusion that the only likely $J^{PC}$ assignments for the $X(3872)$
are $1^{++}$ and $2^{-+}$~\cite{CDF_angles}.  

The unfilled charmonium state near 3872~MeV with $J^{PC}=1^{++}$ is the $2^3P_1$
($\chi_{c1}^{\prime}$).  However, charmonium models predict this state to have a mass
of $\simeq 3905$~MeV, much higher than the world average $X(3872)$ mass, 
$M_{X(3872)} = 3871.56\pm 0.22$~MeV~\cite{PDG}.  The predicted mass is
tightly constrained by the fact that the multiplet partner state,
the $\chi_{c2}^{\prime}$ has been found and its mass
measured to be $3929\pm 6$~MeV~\cite{belle_z3930}.  
The unfilled $\ccbar$ state near 3872~MeV with $2^{-+}$ is the 
 $1^1D_2$ ($\eta_{c2}$) state.  However, the model
prediction for the mass, $\simeq 3837$~MeV, is too low, a prediction
that is also tightly constrained, this time by the measured mass of
the well established $\psi(3770)$~\cite{PDG} multiplet partner state.

A striking feature
of the $X(3872)$ is that its mass is equal within
rather small errors to the $D^0\bar{D^{*0}}$ mass threshold,
$m_{D^0}+m_{D^{*0}}= 3871.79\pm 0.30$~MeV, and this has prompted 
speculation that it is a molecule-like  $D^0\bar{D^{*0}}$ bound 
state~\cite{molecule}.
Deuteron-like interactions between
$D^0$ and $\bar{D^{*0}}$ mesons were studied by 
T\"{o}rnqvist in 1994, and he predicted bound states for 
for $J^{PC}$ values of $0^{-+}$ and $1^{++}$~\cite{tornqvist}. 
Now there is a growing consensus that the $X(3872)$ is a  $1^{++}$ 
$D^0\bar{D^{*0}}$ bound state with some
admixture of the $\chi_{c1}^{\prime}$.  The $\chiconep$
component is supposed to be responsible for its prompt production
in $\ppbar$ collisions and its decay transitions to charmonium states.

\subsection{Radiative transitions of the $X(3872)$}

Important diagnostics for distinguishing between various possibilities
are radiative $X(3872)\rt\gamma\psip$ and $\gamma\jp$ decays.
If the $X(3872)$ is the $\chiconep$ or if it is a mixed
state where the $\chi_{c1}^{\prime}$ component is primarily responsible
for its inter-charmonium transitions, one can expect its
partial width for $X(3872)\rt\gamma\psip$, for which there
is good wave function overlap, to be substantially larger than that
for  $X(3872)\rt\gamma\jp$, which are hindered by the poor match
of the initial \& final-state radial wave functions.  A potential model
calculation indicates that the $\gamma\psip$ transition is favored by
more than a factor of ten~\cite{barnes}.
For the $X(3872)=\eta_{c2}$ case, the situation is reversed and the
$\gamma\jp$ mode is favored by an order-of-magnitude~\cite{jia}.

\begin{figure}[htb]
  \includegraphics[height=0.2\textheight]{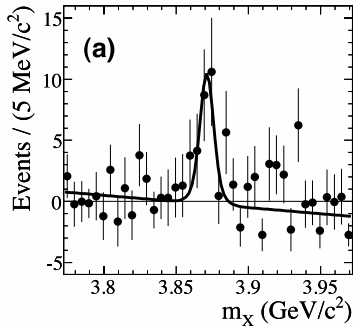}
  \caption{The $\gamma \psip$ invariant mass distribution near
3872~MeV for $B^+\rt K^+\gamma\psip$ decays from BaBar. 
Note that this distribution is background subtracted. }
\label{fig:babar_gammapsip}
\end{figure}

\begin{figure}[htb]
  \includegraphics[height=0.125\textheight]{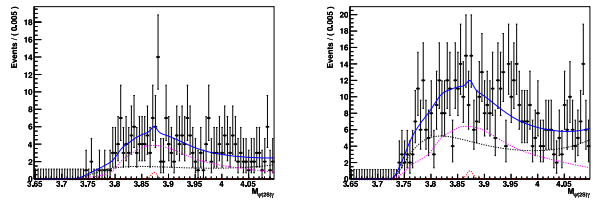}
  \caption{ The $\gamma \psip$ invariant mass distribution near
3872~MeV for $B^+\rt K^+\gamma\psip$ decays from Belle.  The left-side
panel shows results from the sample where $\psip\rt\leplep$, the
right-hand shows the results when $\psip\rt\pipi\jp$.}
\label{fig:belle_gammapsip}
\end{figure}

In 2009, BaBar reported  $>3\sigma$ significance signals for $X(3872)$ decays
to both $\gamma\jp$ and $\gamma\psi'$~\cite{babar_gammajpsi}, 
(see Fig.~\ref{fig:babar_gammapsip}) with the $\gamma\psi'$ decay mode
favored over the  $\gamma\jp$ transition by a factor
of $3.4\pm 1.4$. This year Belle reported preliminary results that
claim a $>5\sigma$ signal for $X(3872)\rt\gamma\jp$
at a rate that agrees with BaBar but saw no evidence for
$X(3872)\rt\gamma\psip$ (see Fig.~\ref{fig:belle_gammapsip}).
Belle set a 90\% CL upper limit
on the $\gamma\psip / \gamma\jp$ ratio of  $<2.1$,
below the BaBar central value~\cite{vishnal_qwg7}.
In any case, the large preference for $\gamma\psip$ compared to
$\gamma\jp$ that is expected for the $\chiconep$
is not seen.

\subsection{$X(3872)\rt\omega \jp$}
\begin{figure}[htb]
  \includegraphics[height=0.125\textheight]{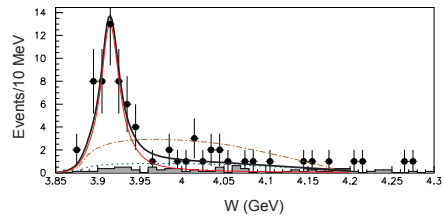}
  \caption{ The CM energy distribution for $\gamma\gamma\rt\omega\jp$
from Belle.}
\label{fig:belle_gg_y3940}
\end{figure}

In 2005, Belle reported  a near-threshold
$\omega J/\psi$ mass peak in the decay $B\rightarrow K\omega J/\psi$
that they called the $Y(3940)$~\cite{belle_y3940}.
The $Y(3940)$ mass is well above open-charm mass thresholds for decays
to $D\bar{D}$ or $D^*\bar{D}$ final states, but was discovered via its decay to
the hidden charm $\omega\jp$ final state.  A search for
$Y(3940)\rt D^*\bar{D}$ decays resulted in a 90\% CL {\it lower} limit on
the ratio ${\mathcal B}( Y(3940)\rt\omega\jp ) 
/ {\mathcal B}(Y(3940)\rt D^0\bar{D^{*0}})>0.71$~\cite{belle_y3940_2_DDstr}.   
This implies an  $\omega\jp$ partial width
that is much larger than expectations for charmonium.  
The $Y(3940)$ sighting in $B\rt K\omega\jp$ decays was 
confirmed by Babar in 2008~\cite{babar_y3940}.
Recently Belle reported a near-threshold $\omega\jp$ mass peak
in the untagged two-photon process $\gamma\gamma\rt\omega\jp$
with resonance parameters $M=3915\pm 4$~MeV and $\Gamma = 17\pm 11$~MeV,
which are consistent with those of the $Y(3940)$ 
(see Fig.~\ref{fig:belle_gg_y3940})~\cite{uehara_y3940}.
If this is the $Y(3940)$, it narrows the $J^{PC}$
quantum numbers down to $0^{\pm +}$ or $2^{\pm +}$.
Belle measures $M=3915\pm 4$~MeV and $\Gamma = 17\pm 11$~MeV
$\Gamma_{\gamma\gamma}{\mathcal B}(Y\rt\omega\jp)=
61\pm 19$~eV (for $J^P=0^+$). If
$\Gamma_{\gamma\gamma}\sim {\mathcal O}$(1~keV),
a value typical for charmonium mesons, then
$\Gamma (Y\rt\omega\jp)\sim {\mathcal O}$(1~MeV),
which is very large for a hadronic inter-charmonium transition.

Although the $X(3872)$ mass is below the threshold for 
$X(3872)\rt\omega\jp$ decays, Swanson proposed a composite
model in which the $X(3872)$ has a large $\omega\jp$ component
and that $\omega\jp$ decays to the low-mass tail of the $\omega$ would
be comparable in rate to $\pipi\jp$ decays~\cite{swanson}.
Belle, in a 2005 unpublished paper, reported evidence for subthreshold
$\omega\jp$ decays at a rate comparable to that for $\pipi\jp$, consistent
with the Swanson prediction~\cite{belle_X_2_omega_jpsi}.
This year, the BaBar group reported evidence for $X(3872)\rt\omega\jp$~\cite{babar_x_2_omega-jpsi}
at a rate consistent with that reported by Belle and the Swanson prediction.

A BaBar fit to the $\pipi\pi^0$ lineshape for the selected 
$X(3872)\rt\omega\jp$ events that assumed an odd parity
for the $X(3872)$ had a better $\chi^2$ value than a
fit that asuumed even parity: $\chi^2/d.o.f. = 3.53/5$
for odd parity  as opposed to $\chi^2/d.o.f. = 10.17/5$ for
even parity.   While the statistical significance of this 
difference is not overwhelming  (less than $2\sigma$),
it has led to some reconsideration of the $2^{-+}$ 
assignment~\cite{yulia}.

\begin{figure}[htb]
  \includegraphics[height=0.12\textheight]{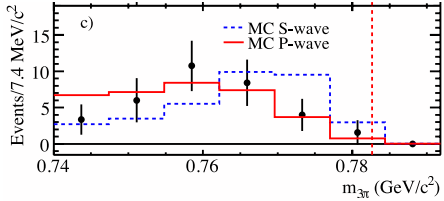}
  \caption{BaBar's $\pipi\pi^0$ invariant mass distribution for
$X(3872)\rt\pipi\pi^0 \jp$ decays. The solid (dashed) histogram shows
results from an odd (even) parity fit. }
\label{fig:babar_omega-jpsi}
\end{figure}

\section{Recently reported $ \phi\jp$ peaks}

\begin{figure}[htb]
  \includegraphics[height=0.15\textheight]{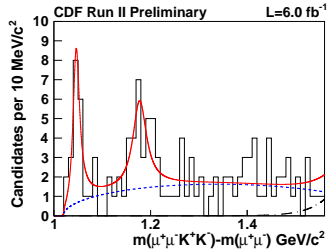}
  \caption{  The $M(\phi \leplep) -M(\leplep )$  distribution 
for $B^+\rt K^+ \phi \leplep$ decays where $M(\leplep)$ is
in the $\jp$ region from CDF.  The peak at threshold is the $Y(4140)$,
the second peak is at a mass of $\simeq 4275$~MeV.}
\label{fig:y4140_2010}
\end{figure}

In 2009, CDF reported a narrow 
$14\pm 5$ event near-threshold peak in
the $M(\phi\jp)$  distribution from 
$B\rt K\phi\jp$ decays~\cite{CDF_y4140}.
This summer, they reported an update with about twice the data
where the excess has grown to a $19\pm 6$ signal with
a $5.9\sigma$ statistical significance (see Fig.~\ref{fig:y4140_2010}). 
The mass and width from the larger sample, 
$M=4144\pm 3$~MeV and $\Gamma = 15^{+10}_{-6}$~MeV, 
agree well with previous results~\cite{CDF_y4140_v2}.
They also report hints of a higher mass peak at $\simeq 4275$~MeV
but with marginal significance.  The mass of the $Y(4140)$ is well
above all open-charm thresholds and, thus, such a narrow peak with a strong
$\phi\jpsi$ component is not expected for an ordinary $\ccbar$ 
meson.   The similarities between the
$Y(3940)\rt \omega\jp$ seen by Belle \& BaBar and the CDF
group's $Y(4140)\rt\phi\jp$ suggests that they may
originate from related sources~\cite{molina}.
 
\begin{figure}[htb]
  \includegraphics[height=0.15\textheight]{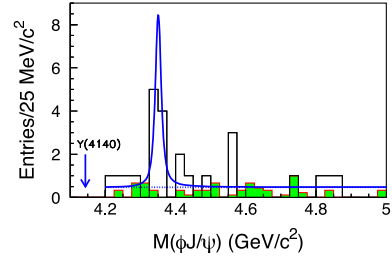}
  \caption{  The $X(4350)$ peak in the $M(\phi \jp)$  distribution 
for $\gamma\gamma \rt \phi \jp$ events 
from Belle.}
\label{fig:belle_x4350}
\end{figure}

At the $B$-factories, the $B$ mesons are produced nearly
at rest. Thus, in the process $B\rt K Y(4140)$, $Y(4140)\rt \phi\jp$
the kaons from $\phi\rt K^+K^-$ have very low momentum and
a very small detection efficiency.  As a result, neither Belle
nor BaBar have been able to either confirm or
contradict the CDF observation.  On the
other hand, Belle studied $\phi\jp$ systems produced via 
$\gamma\gamma\rt\phi\jp$.  They did not see the $Y(4140)$, but did
see evidence (with $3.8\sigma$ statistical significance) for a narrow peak 
that they dubbed the $X(4350)$ with mass
$4350\pm 5$~MeV and width $\Gamma=13^{+18}_{-9}$~MeV~\cite{belle_x4350}
(see  Fig.~\ref{fig:belle_x4350}).   An
interesting spectroscopy  in the $\phi\jp$
channel seems to be emerging.

\section{Threshold $M(\ppbar)$ peak in $\jp\rt\gamma\ppbar$}

In 2003, BESII reported the observation of a
striking enhancement in the $M(\ppbar)$ distribution in
radiative $\jp\rt\gamma\ppbar$ decays~\cite{bes2_ppb}.  
The result of a fit to a Breit Wigner shape was a peak mass
of $1859^{+3}_{-10}$~MeV, about 18~MeV below the $M(\ppbar)=2m_p$
mass threshold, and a width $\Gamma<30$~MeV (90\% CL).
These parameters do not match those of any known resonance.
Similar enhancements are not seen
in $\psip$ or $\Upsilon\rt\gamma\ppbar$ or $\jp\rt\omega\ppbar$ and
the enhancement cannot be fit with $\ppbar$ final state interactions.  

Ding and Yan suggested that this might be a bound
$\ppbar$ state (baryonium) in which case it
might also be seen to decay to $\pipi\eta^{\prime}$~\cite{ding}. 
In a study of $\jp\rt\gamma\pipi\eta^{\prime}$,
BESII found a $\pipi\eta^{\prime}$ mass 
peak at $1834\pm 7$~MeV with width
$\Gamma =68\pm 21$~MeV (the $X(1835)$)~\cite{bes2_x1835}.  It
is not clear if the $X(1835)$ is related to the $\ppbar$ peak.

\begin{figure}[h!]
  \includegraphics[height=0.15\textheight]{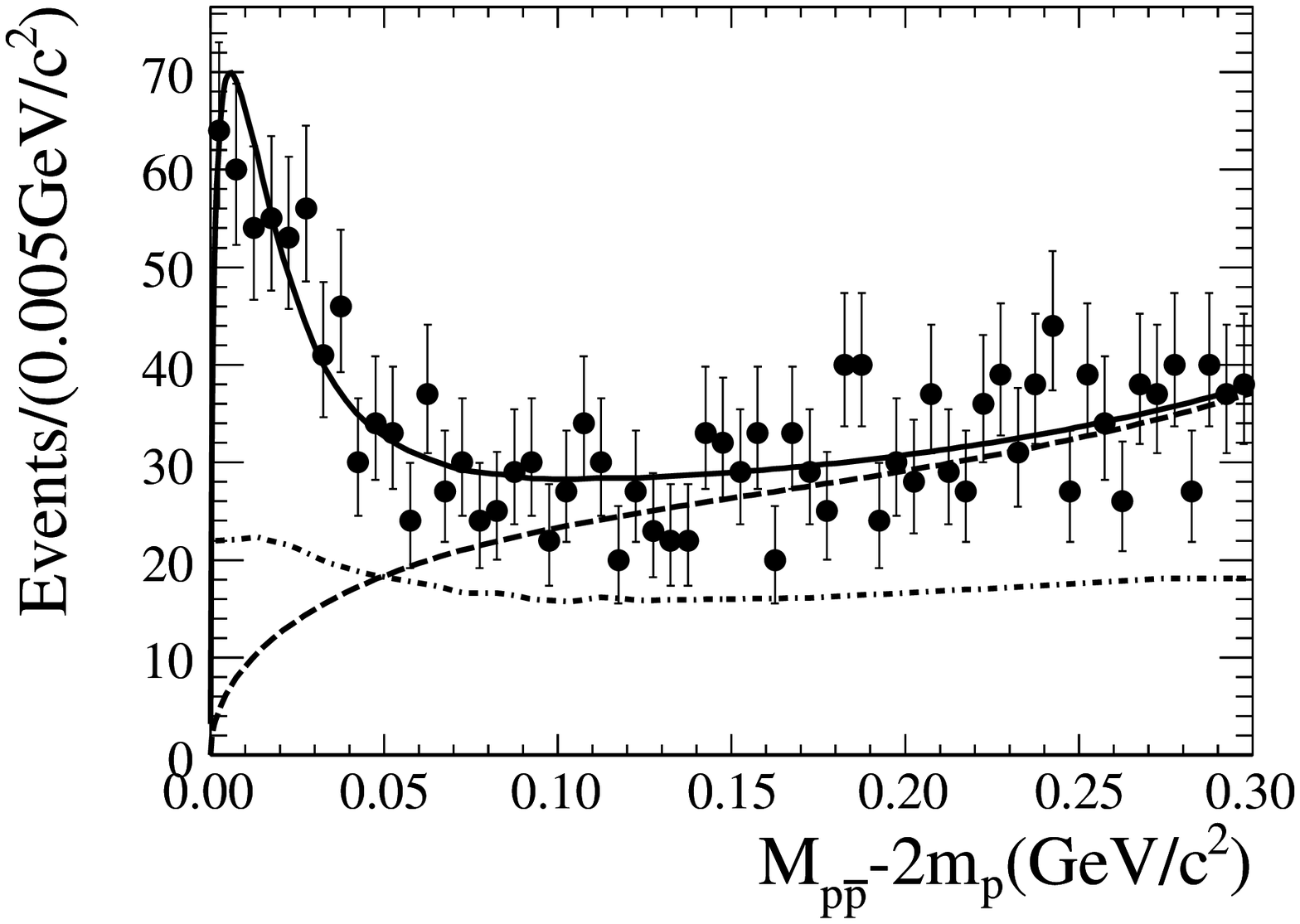}
  \caption{  The $\ppbar$ invariant mass distribution 
for $\psip\rt \pipi \jp$, $\jp\rt \gamma\ppbar$ events 
from BESIII.}
\label{fig:bes3_ppb}
\end{figure}

An early task at the  BESIII experiment
has been the confirmation of the above-mentioned observations.
Figure~\ref{fig:bes3_ppb} shows the $M(\ppbar )$ distribution
from $\jp\rt\gamma\ppbar$ decays for $\jp$s produced
via $\psip\rt\pipi\jp$ decays in a 108M $\psip$ event sample~\cite{bes3_ppb}.
The threshold enhancement is
evident; a fit gives $M=1861^{+6}_{-13}$~MeV
and $\Gamma < 38$~MeV, consistent with the BESII
results.

\begin{figure}
  \includegraphics[height=0.2\textheight]{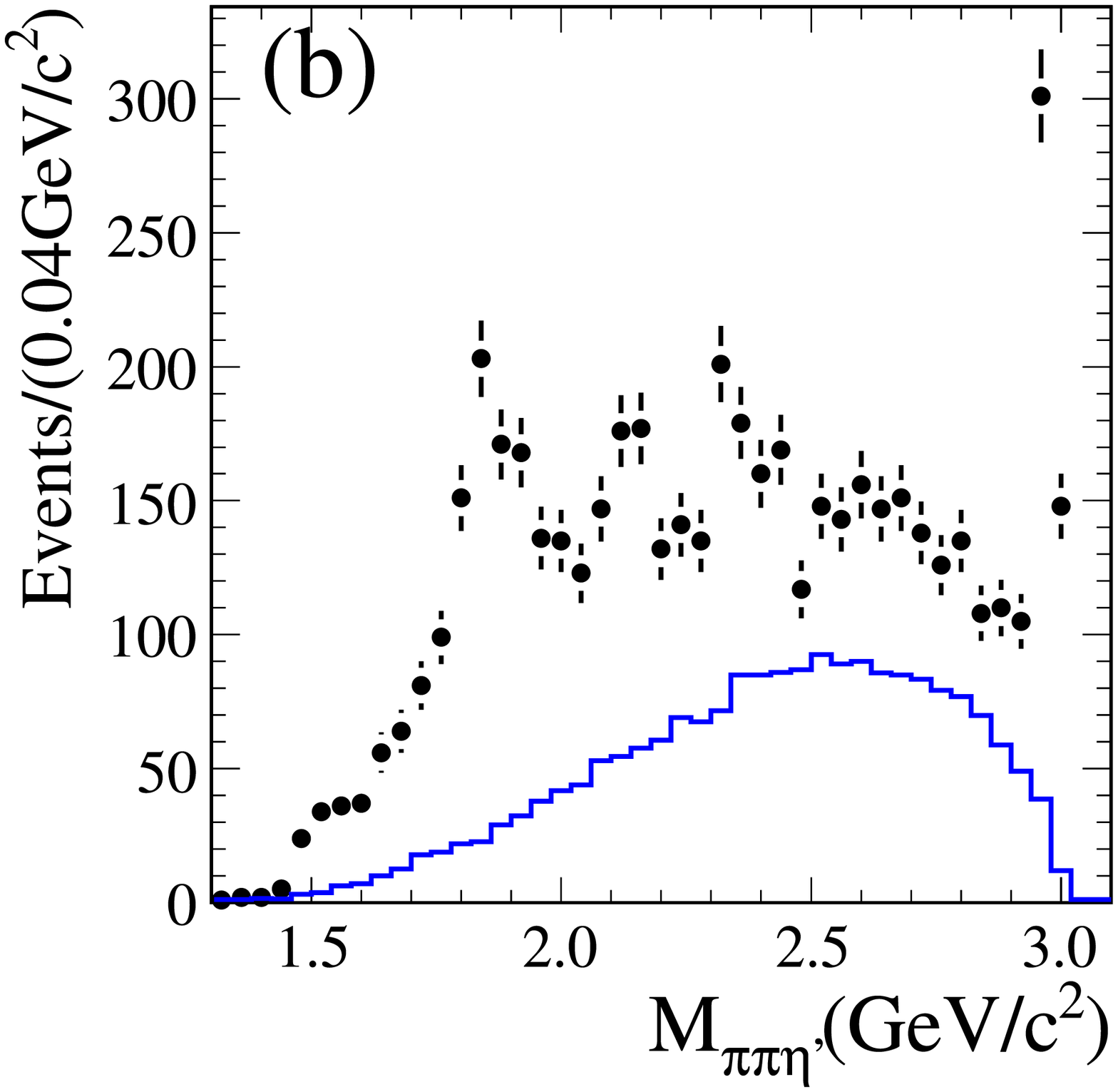}
  \caption{  The $\pipi\eta^{\prime}$ invariant mass distribution 
for $\jp\rt \gamma\pipi\eta^{\prime}$ events 
from BESIII.}
\label{fig:bes3_x1835}
\end{figure}

BESIII also studied the $\jp\rt\gamma\pipi\eta^{\prime}$ process
with a 226M $\jp$ event sample.  The resulting $M(\pipi\eta^{\prime})$
distribution is shown in Fig.~\ref{fig:bes3_x1835}. In addition
to a prominent $X(1835)$ signal, two other peaks are evident at
higher masses, as well as a large $\eta_c\rt\pipi\eta^{\prime}$ signal
near~3.0 GeV. 
Preliminary BESIII results
for the mass and width for the $X(1835)$
 were reported this summer: 
$M=1836.5\pm 3.0 {\rm (stat)}^{+5.6}_{-2.1} {\rm (syst)}$~MeV
and $\Gamma = 190\pm 9 {\rm (stat)} ^{+31}_{-36} {\rm (syst)}$~MeV~\cite{bes3_x1835}.
The mass agrees well with the BESII result while the width is considerably
broader.  The BESIII results confirm those from BESII, but the discrepancy
between the width values for the $\ppbar$ and $\pipi\eta^{\prime}$ peaks
has increased, making it less likely that the two structures
are related.  The $X(1835)$ and its higher mass partners may be excitations
of the $\eta^{\prime}$.

\section{$\sigma (\ee\rt\pipi\Upsilon(nS))$ at 10.9~GeV}

Perhaps the most mysterious of the $XYZ$ mesons are the $1^{--}$ 
$Y(4260)\rt\pipi\jp$, $Y(4350)\rt\pipi\psip$ and $Y(4660)\rt\pipi\psip$,
first found by BaBar~\cite{babar_y4260,babar_y4325} and confirmed
by Belle~\cite{belle_y4260,belle_y4325} in the initial-state-radiation
process $\ee\rt\gamma_{isr}\pipi\jp(\psip)$.  These states have
much larger partial widths to $\pipi\jp$ ($\pipi\psip$) than
those for $\psip\rt\pipi\jp$ ($102\pm 3$~keV) or 
$\psi(3770)\rt\pipi\jp$ ($53\pm7$~keV). In fact, the $Y(4260)$ mass
coincides with a dip in the $\ee\rt hadrons$ total cross 
section~\cite{bes_Rhad} and it has a full width of $95\pm 14$~MeV~\cite{PDG}.
This implies a {\it lower} limit on its $\pipi\jp$ partial
width of $\sim 1$~MeV~\cite{moxh}.  (Since no other decay modes
have yet been identified, the $\pipi\jp$ partial width may be much
larger.)  This motivated the Belle experiment to look
for similar phenomena in the $b$-quark sector~\cite{weishu}.

\begin{figure}
  \includegraphics[height=0.15\textheight]{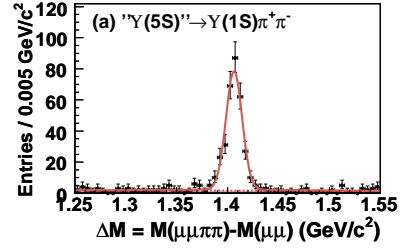}
  \caption{  The $M(\pipi\mumu)-M(\mumu)$ distribution 
for $\Upsilon(5S)\rt \pipi\mumu$ events with $M(\mumu)$
in the $\Upsilon(1S)$ mass range from Belle.}
\label{fig:belle_Y5S_2_Y1S}
\end{figure}

Using their huge sample of 464 million $\Upsilon(4S)$ decays 
(accumulated for making measurements of $CP$ violation in
$B$ meson decays), Belle detected $113\pm 16$ events of
the type $\Upsilon(4S)\rt\pipi \Upsilon(1S)$ ($\Upsilon(1S)\rt\mumu$),
from which it determined the partial width to
be $\Gamma(\Upsilon(4S)\rt\pipi\Upsilon(1S))=3.65\pm 0.95$~keV~\cite{belle_Y4S},
in agreement with expectations for bottomonium mesons.
In 2008, Belle accumulated a much smaller sample of 6.5 million $\Upsilon(5S)$ for
pilot studies of $B_s$ decays. According to standard bottomonium
expectations normalized by the $\Upsilon(4S)$ measurements, this small
sample of events should contain at most one or two 
$\Upsilon(5S)\rt\pipi\Upsilon(1S)$ events.  Instead, Belle
observed the distinct $325\pm 20$ event signal shown 
in Fig.\ref{fig:belle_Y5S_2_Y1S}.  A similarly distinct $186\pm 15$ 
event signal was seen for $''\Upsilon(5S)''\rt\pipi\Upsilon(2S)$. 
(I use inverted commas to emphasize that it is not
known that the $\Upsilon(5S)$ is in fact the source for these events.)
Assuming these signals are from the $\Upsilon(5S)$,
Belle infers partial widths of
$590 \pm 10$~keV and $850\pm 18$~keV for the
$\pipi\Upsilon(1S)$ and $\pipi\Upsilon(2S)$ transitions,
respectively, both of which are more than 100~times expectations~\cite{belle_Y5S_2_Y1S}.

\begin{figure}
  \includegraphics[height=0.15\textheight]{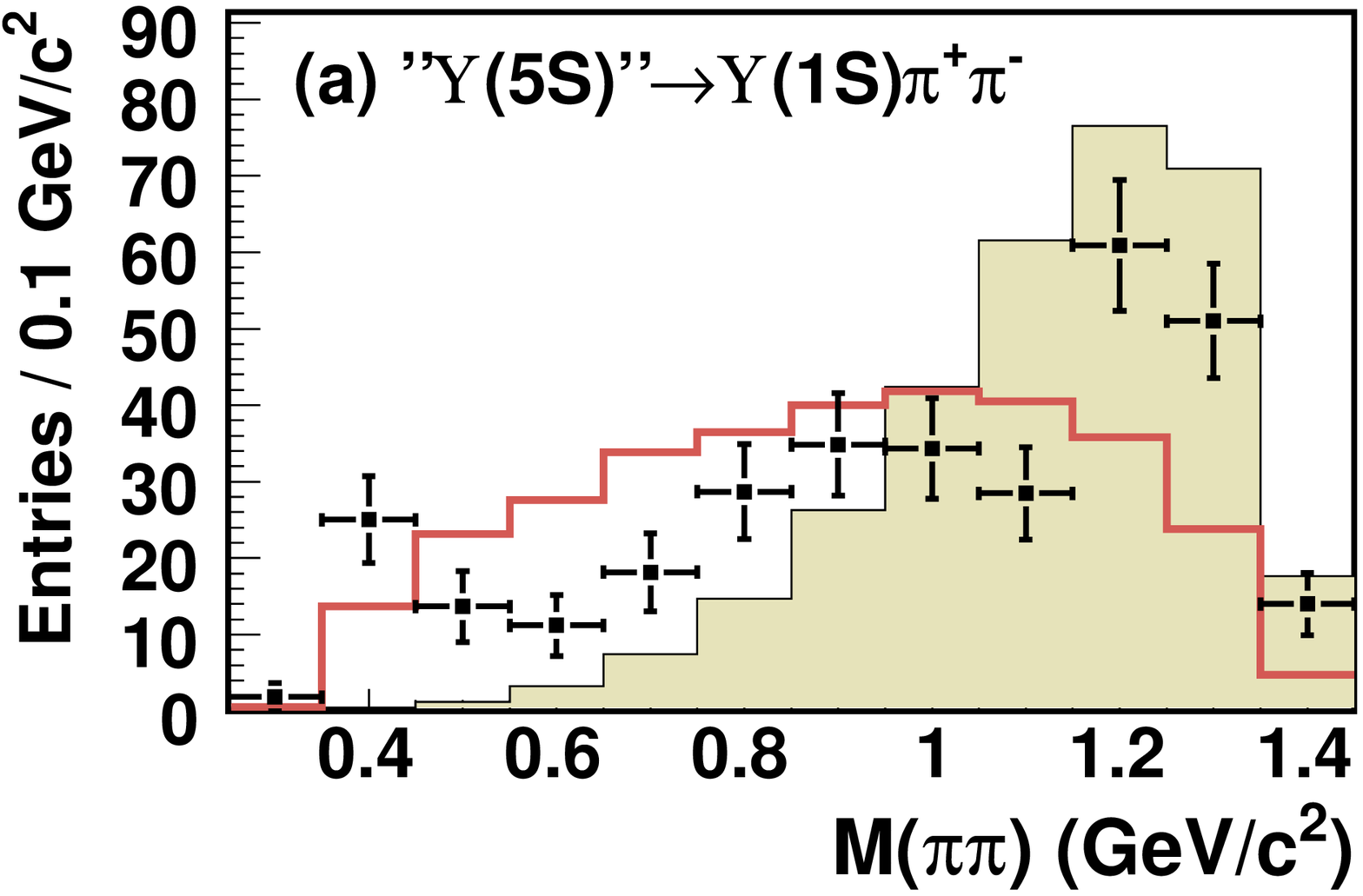}
  \caption{  The $M(\pipi)$ distribution 
for the  $''\Upsilon(5S)''\rt\pipi\Upsilon(1S)$ events from Belle.}
\label{fig:belle_mpipi_Y1S}
\end{figure}

An important question is whether or not the source of these
anomalous $\pipi\Upsilon(nS)$ events is the
$\Upsilon(5S)$, enhanced by some dynamical process,
or if they are they from a $b$-quark sector equivalent to
the $Y(4260)$.    Meng and Chao explored the former approach 
and proposed a model that attributed the anomalous $\pipi\Upsilon(1S)$
and $\pipi\Upsilon(2S)$ production rates at the $\Upsilon(5S)$
to rescattering processes of the type 
$\Upsilon(5S)\rt B^{(*)}\bar{B^{(*)}}\rt f\Upsilon(nS)$, where
$f$ denotes scalar $\pipi$ resonances such as the $\sigma$, the
$f_0(980)$ and/or the $f_0(1370)$~\cite{chao_1}.
However, their approach has some problems.  First, in their
model about two thirds of the contribution to the 
$\pipi\Upsilon(1S)$ channel is due to the $f_0(980)$.  However,
the measured $M(\pipi)$ spectrum for this process from
ref.~\cite{belle_Y5S_2_Y1S}, shown in Fig.~\ref{fig:belle_mpipi_Y1S},
shows no sign of a significant $f_0(980)$ contribution. 

\begin{figure}
  \includegraphics[height=0.15\textheight]{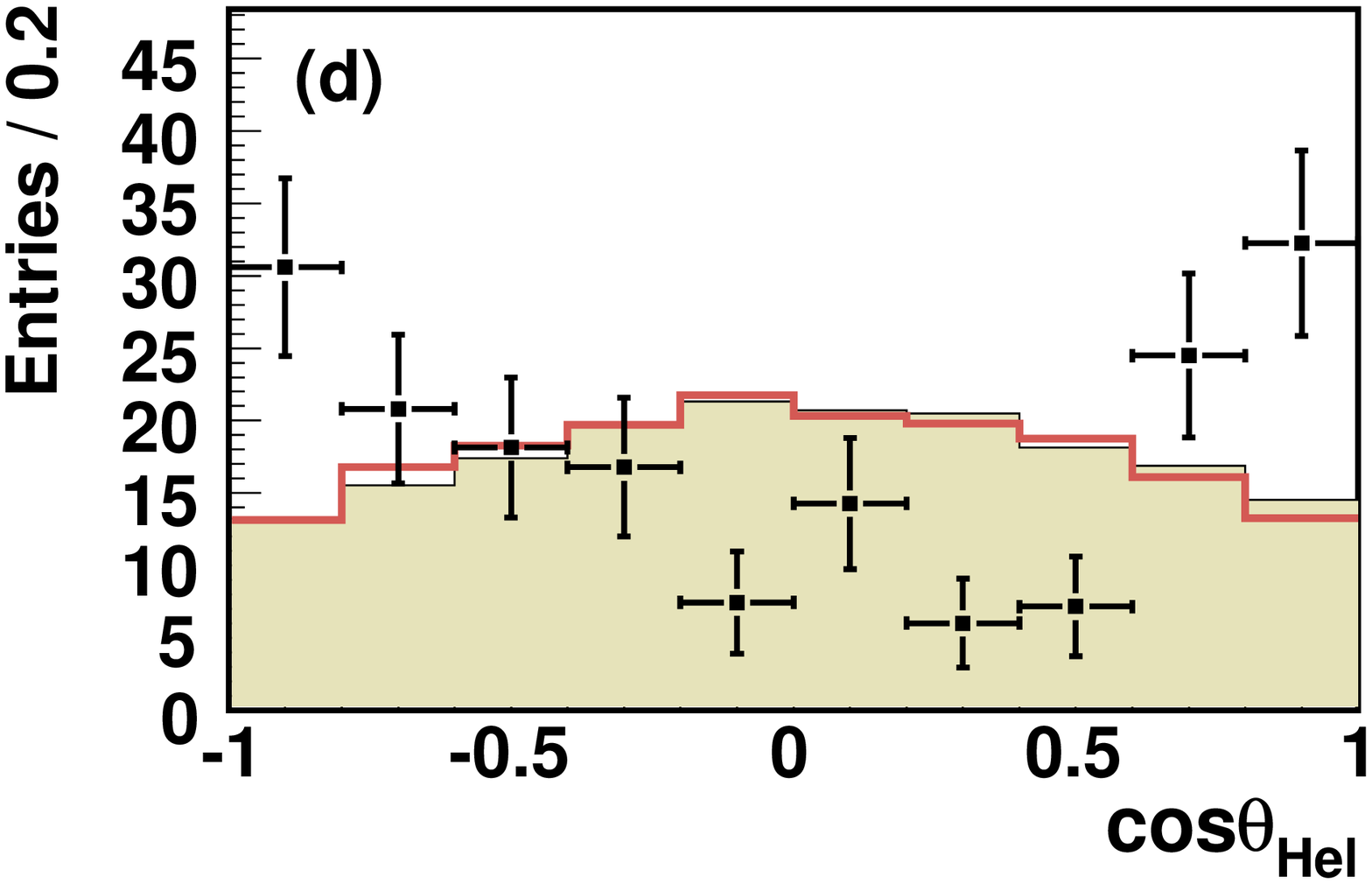}
  \caption{  The $\cos\theta_{\rm Hel}$ distribution 
for the  $''\Upsilon(5S) ''\rt\pipi\Upsilon(2S)$ events from Belle.}
\label{fig:belle_costhel_Y2S}
\end{figure}

A second difficulty with the model can be seen in 
Fig.~\ref{fig:belle_costhel_Y2S}, also from 
ref.~\cite{belle_Y5S_2_Y1S}, where the data points show the
$\cos\theta_{\rm Hel}$ distribution for the $\pipi$ system 
in the $''\Upsilon(5S) ''\rt\pipi\Upsilon(2S)$ events,
where $\theta_{\rm Hel}$ is the angle between the $\pi^+$
and the $\pipi$ system boost direction in the $\pipi$ CM.
Here large and significant deviations from an acceptance-weighted
flat distribution (indicated by the histograms) are evident,
contrary to expectations for $S$-wave $\pipi$ systems. 

\begin{figure}
\includegraphics[height=0.13\textheight]{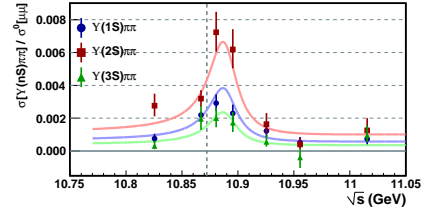}
\caption{  The $\sqrt{s}$ dependence of 
$\sigma(\ee\rt\pipi\Upsilon(nS))$ for $n=1$ (circles)
$n=2$ (squares) and $n=3$ (triangles) from Belle.
The curves are the result of the fit described
in the text and the dashed vertical line indicates the
$\Upsilon(5S)$ peak position.}
\label{fig:belle_pipiY1S_scan}
\end{figure}

If the anomalous $\pipi\Upsilon(nS)$
events are due to an $Y(4260)$-like particle in the $b$-quark
sector, their peak mass and total width
would not necessarily
coincide with the corresponding $\Upsilon(5S)$ parameters.
Belle investigated this with an energy scan around
the $\Upsilon(5S)$ peak that measured the $\sqrt{s}$ dependence
of $\pipi\Upsilon(nS)$ production ($n=1,~2~\&~3)$.  The results
of the scan, shown in Fig.~\ref{fig:belle_pipiY1S_scan}, are that
these event have a peaking structure and that the
peak mass and full width, determined from a single BW fit to the three
channels simultaneously are $M=10889^{+6}_{-3}$~MeV and 
$\Gamma = 37^{+16}_{-10}$~MeV~\cite{belle_y10890}.
The measured peak mass value differs from the recent precise measurement
by BaBar of  $M_{\Upsilon(5S)} =  100876\pm 2$~MeV -- indicated by
the vertical dashed line in  Fig.~\ref{fig:belle_pipiY1S_scan} -- by $2\sigma$ 
(systematic effects included)~\cite{babar_Y5S}. 
BaBar measures $\Gamma_{\Upsilon(5S)} =  43\pm 4$~MeV, which is
narrower than the PDG value and is not distinct from Belle's fitted width 
of the $\pipi\Upsilon(nS)$ peak.  

\begin{figure}[t!]
\centering
\vbox{
\includegraphics[width=6.1cm,height=2.10cm]{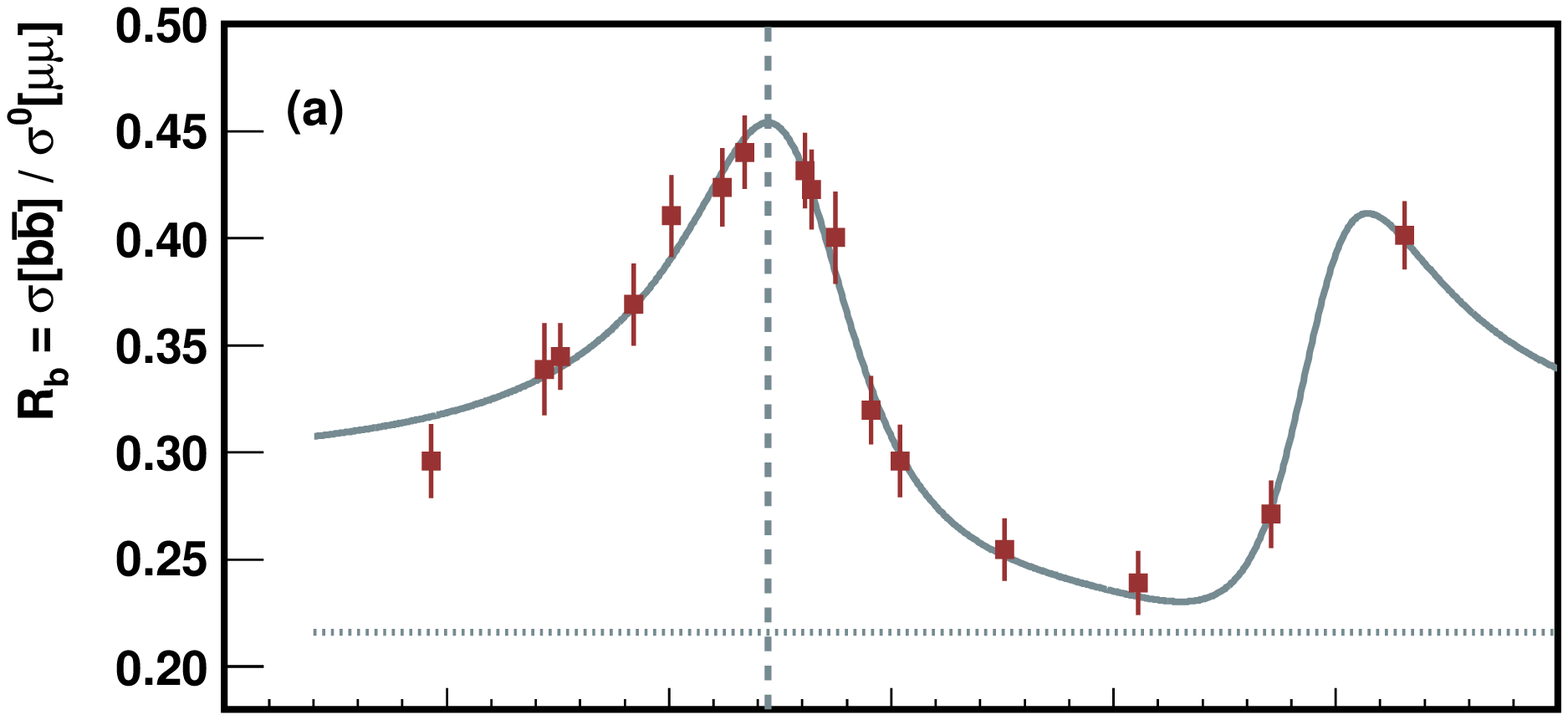}
\includegraphics[width=6.1cm,height=2.10cm]{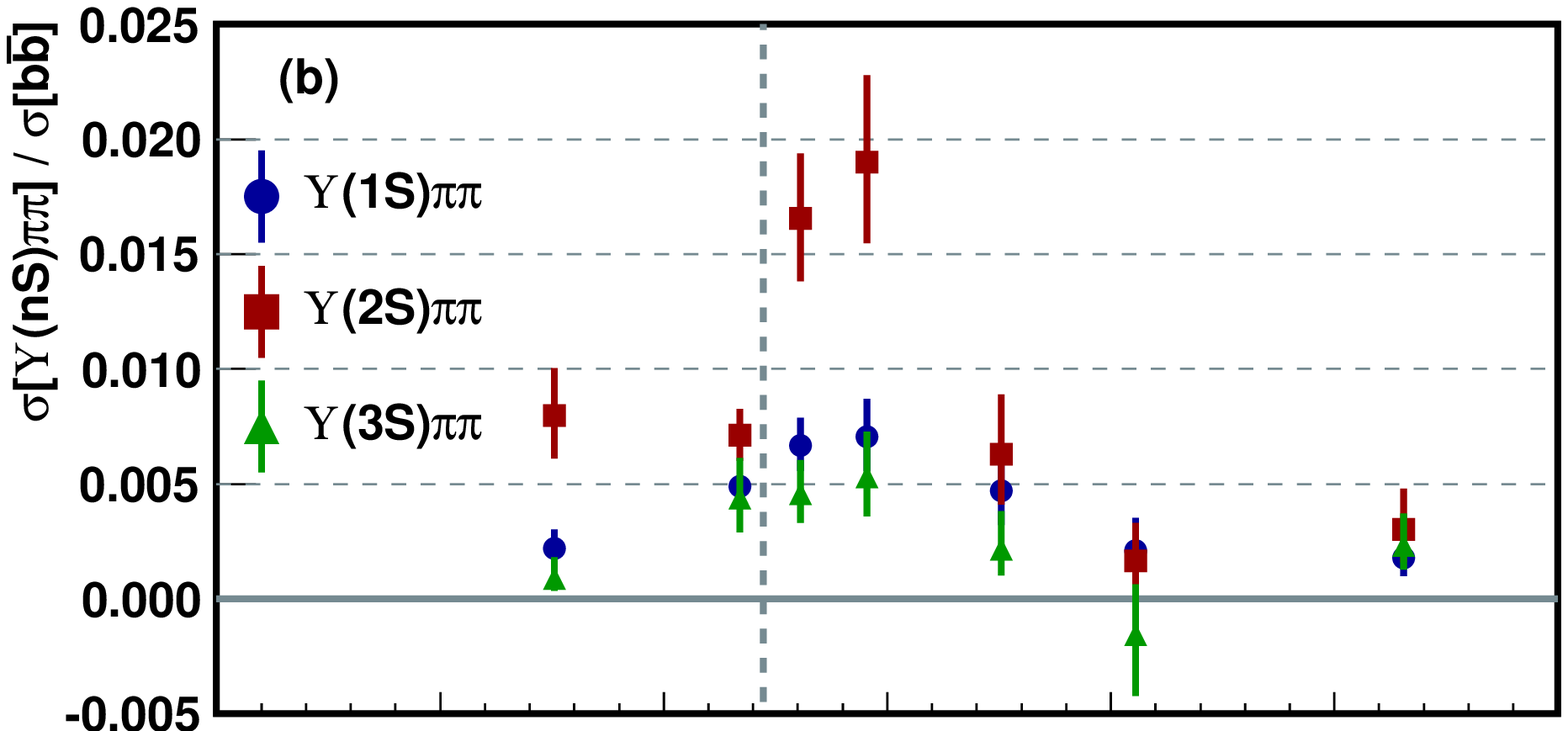}
\includegraphics[width=6.1cm,height=2.52cm]{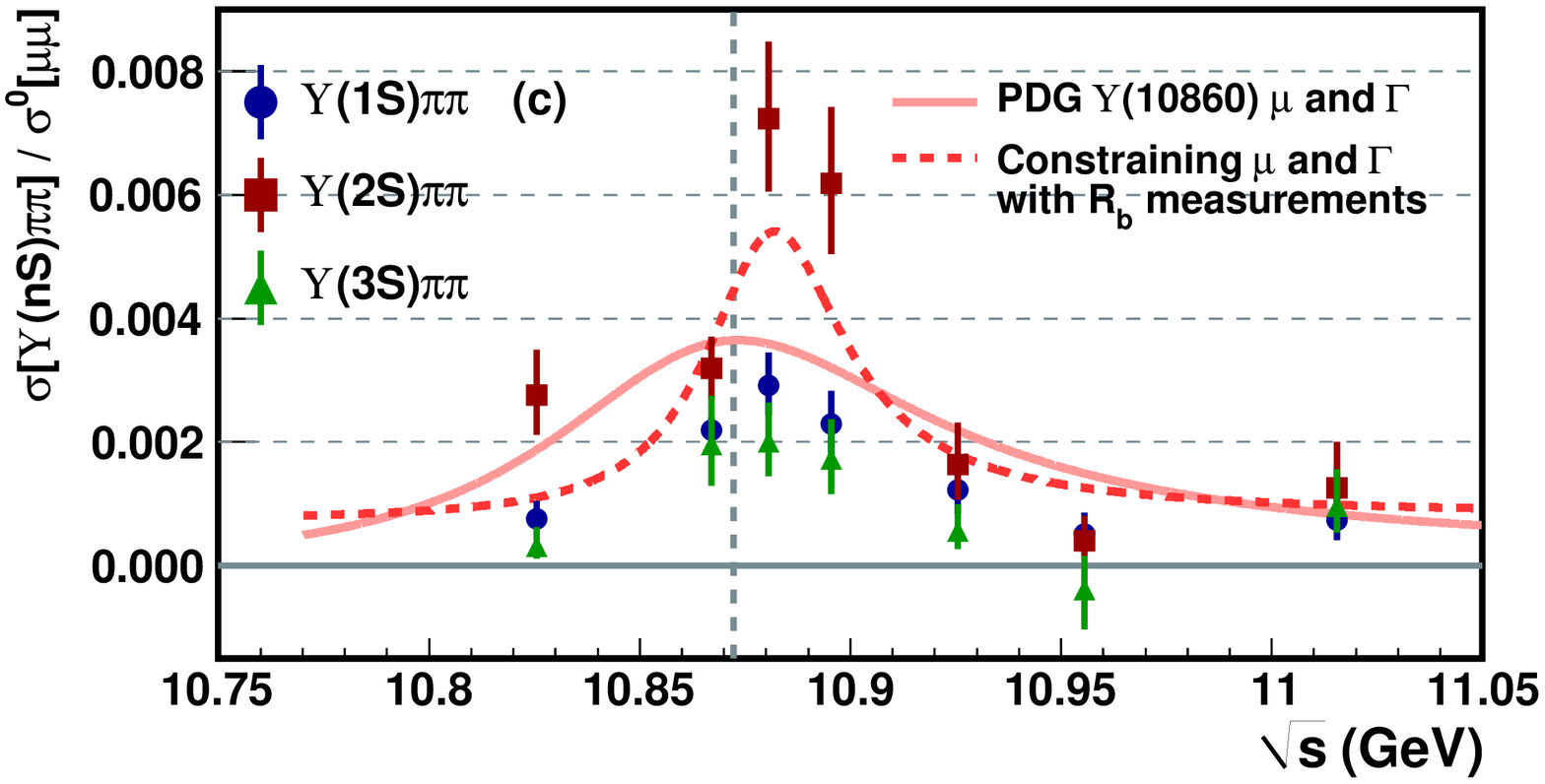}
}
\caption{(a) $R_b$ and (b) $\sigma(e^+e^-\rt \pipi\Upsilon(nS))/\sigma(e^+e^-\rt b\bar{b})$ 
(c) $R_{\pipi\Upsilon(nS)}$, 
the results of fits with resonance parameters from the $R_b$ fit \& the PDG are super
imposed. .The vertical dashed line indicates the $\sqrt{s}$ value where $R_{b\bar{b}}$
is maximum.  From ref.~\cite{belle_y10890}.}
\label{fig:had_ratio}
\end{figure}

The situation is summarized in Fig.~\ref{fig:had_ratio}, where Belle
measurements of $R_{b\bar{b}}$, the total cross secton for 
$\ee \rt b\bar{b}$ normalized to $\sigma_{0}(\mumu)$ is shown
in  Fig.~\ref{fig:had_ratio}a, 
the ratio of $\sigma (\ee\rt \pipi\Upsilon(nS))/\sigma(\ee\rt b\bar{b})$
is shown in  Fig.~\ref{fig:had_ratio}b, and $R_{\pipi\Upsilon(nS)}$
in  Fig.~\ref{fig:had_ratio}c.  In the top figure the curve is
the result of a fit with the $\Upsilon(5S)$ and
$\Upsilon(6S)$ and an interfering non-resonant background (dashed horizontal line).  
The $\Upsilon(5S)$ parameters are allowed to float, the $\Upsilon(6S)$ parameters
are taken fixed at the ref.~\cite{babar_Y5S} values.  The dashed curve in the bottom
figure is the fit to $\pipi\Upsilon(1S)$ data with mass and width constrained by 
the $R_{b\bar{b}}$ fit.  When the $R_{b\bar{b}}$ constraint is relaxed, the $\chi^2$
reduces by 8.71 with an increase of two degrees of freedom, indicating a $\sim 2.5\sigma$
preference for different parameters for the $b\bar{b}$ and $\pipi\Upsilon(nS)$ peaks.  
The vertical dashed line indicates the position of
the maximum value of $R_{b\bar{b}}$. (The peak in $R_{b\bar{b}}$ is shifted from
the fitted $\Upsilon(5S)$ resonance mass because of interference effects.)

Thus, as opposed to the $Y(4260)$ and the charmonium resonances where
the anomalous $Y(4260)\rt\pipi\jp$ peak occurs at a dip in the
$\ee\rt hadrons$ cross section and far from the masses of the known
$1^{--}$$\ccbar$ resonances,  in this case the data favor
the interpretation that the peak of the anomalous
$\pipi\Upsilon(nS)$ signal is distinct from that
of the $\Upsilon(5S)$ but only at the $\sim 2\sigma$ confidence level.  
Considerably more scanning data is needed
to establish conclusively whether or not the $\Upsilon(5S)$ is the
source of the anomalous events. 
Unfortunately, this will probably not be available at least until
BelleII starts to operate in 2014~\cite{belle2}.

\section{Summary}

Experimental progress on the $XYZ$ particles is reviewed.
Belle and BaBar both see significant signals fof
$X(3872)\rt\gamma\jp$, but 
recent Belle results indicate that the
${\mathcal B}(X(3872)\rt\gamma\psip)$ is not as large as
reported earlier by BaBar.  BaBar confirmed the Belle
sighting of the $Y(3040)\rt\omega\jp$ in $B$ decays
and Belle sees a similar peak in $\gamma\gamma\rt\omega\jp$,
suggesting that its $J^{PC}$ quantum numbers are $0^{\pm +}$
or $2^{\pm +}$.  BaBar confirms the existence of the
subthreshold decay $X(3872)\rt\omega\jp$ and their
fit to the $\pipi\pi^0$ line shape mildly favors
a  $2^{-+}$ over a $1^{++}$ assignment for the $X(3872)$.
CDF confirms their $Y(4140)\rt\phi\jp$ signal with more data
and see hints of another $\phi\jp$ mass peak around 4275~MeV.
Belle sees a different narrow $\phi\jp$ mass peak in $\gamma\gamma$ collisions at
4350~MeV.  BESIII confirms the BESII observations of the
threshhold $\ppbar$ mass peak in $\jp\rt\gamma\ppbar$ decays
and the $X(1835)\pipi\eta^{\prime}$ in $\jp\rt\gamma\pipi\eta^{\prime}$
decays.  They also see two higher mass peaks in the $\pipi\eta^{\prime}$
channel.

The Belle group's discovery of huge
partial widths for $''\Upsilon(5S)''\rt\pipi\Upsilon(nS)$
($n=1~\&~2$) is reviewed.  
Attempts to explain this as a rescattering effect
run into problems with the experimentally measured $M(\pipi)$ and
$\pipi$ helicity angle distributions. Belle measurements of the 
$\sqrt{s}$ dependence of $\sigma(\ee\rt\pipi\Upsilon(nS))$
favor an alternative source for the anomalous events but 
with limited statistical confidence.


\begin{theacknowledgments}

I thank the QCHS-IX organizers for inviting me to give this
talk.  I also thank Yanping Huang, Fred Harris, Kai-Feng Chen 
and Guillermo Rios for their help in preparing this manuscript. 
 This work is supported
by the WCU program of the Ministry of Education Science and Technology
National Research Foundation of Korea.

\end{theacknowledgments}





\begin{thebibliography}{9}
\bibitem{conj} The inclusion of charge-conjugate modes is always implied.
Also, when results are quoted, the reported statistical and systematic errors
are added in quadrature.
\end{thebibliography}
\end{document}